\documentclass[conference]{IEEEtran}
\ifCLASSINFOpdf
\else
\fi

\usepackage{hyperref}
\usepackage{fancyhdr} 
\fancypagestyle{firstpage}{
  \fancyhf{}
  \fancyhead[C]{Pre-print of Best Paper Award IEEE Intelligence and Security Informatics (ISI) 2016 Manuscript}
  \fancyfoot[C]{-~\thepage~-}
}
\usepackage{amssymb}
\usepackage{multirow}
\usepackage{graphicx}
\usepackage{amsmath}
\usepackage[table]{xcolor}
\definecolor{lightgray}{gray}{0.9}
\usepackage{filecontents}
\usepackage{subfig}
\usepackage{array}
\usepackage{amsmath}
\usepackage{pbox}
\usepackage[linesnumbered,ruled]{myalg}
\usepackage{float}
\usepackage{algorithmic}

\usepackage{todonotes}

\renewcommand{\AlCapSty}[1]{\normalfont\scriptsize{\textbf{#1}}\unskip}

\begin{document}

\graphicspath{ {./} }

%
\title{Automated Big Text Security Classification}

\author{\IEEEauthorblockN{Khudran Alzhrani\IEEEauthorrefmark{1},
Ethan M. Rudd\IEEEauthorrefmark{2}, Terrance E. Boult\IEEEauthorrefmark{2} and
C. Edward Chow\IEEEauthorrefmark{1}}
\IEEEauthorblockA{
  University of Colorado at Colorado Springs\\
  \IEEEauthorrefmark{1}\IEEEauthorrefmark{2}Department of Computer Science\\
  \IEEEauthorrefmark{2}Vision and Security Technology (VAST) Lab\\
  Email: \IEEEauthorrefmark{1} \{kalzhran,cchow\}@uccs.edu \IEEEauthorrefmark{2} \{erudd,tboult\}@vast.uccs.edu
}
}


%


\maketitle
\thispagestyle{firstpage}

\begin{abstract}
In recent years, traditional cybersecurity safeguards have proven ineffective against insider threats. Famous cases of sensitive information leaks caused by insiders, including the WikiLeaks release of diplomatic cables and the Edward Snowden incident, have greatly harmed the U.S. government's relationship with other governments and with its own citizens. Data Leak Prevention (DLP) is a solution for detecting and preventing information leaks from within an organization's network. However, state-of-art DLP detection models are only able to detect very limited types of sensitive information, and research in the field has been hindered due to the lack of available sensitive texts. Many researchers have focused on document-based detection with artificially labeled ``confidential documents'' for which security labels are assigned to the entire document, when in reality only a portion of the document is sensitive. This type of whole-document based security labeling increases the chances of preventing authorized users from accessing non-sensitive information within sensitive documents. In this paper, we introduce Automated Classification Enabled by Security Similarity (ACESS), a new and innovative detection model that penetrates the complexity of big text security classification/detection. To analyze the ACESS system, we constructed a novel dataset, containing formerly classified paragraphs from diplomatic cables made public by the WikiLeaks organization. To our knowledge this  paper is the first to analyze a dataset that contains actual formerly sensitive information annotated at paragraph granularity.
\end{abstract}


%
\IEEEpeerreviewmaketitle

\section{Introduction}
Drawing a line between security and convenience that separates users' rights from the misuse of data has never been without complication. 
A common industrial and academic approach is to quantify and limit the amount of sensitive information revealed to insider threats -- i.e., legitimate but potentially malicious users\cite{chen2011auditing,harel2010m}.
However, this approach does not scale well with information explosion, which makes the information leak prevention problem considerably more difficult. 
Data sizes are  expanding exceptionally every year. Gantz et al. \cite{gantz2012digital}  stated that more than 2.8 trillion gigabytes of data were generated in 2012 alone. 
Undoubtedly, among the vast amount of newly generated data, some sensitive information will simply go unlabeled, and a malicious or unaware user might leak unlabeled, sensitive information without being noticed. 
Concerning unstructured textual data, documents may or may not consist of a mixture of sensitive and non-sensitive contents. 
Thus some form of autonomous paragraph-based security labeling is needed to close the gap between security requirements and data availability. 

Sensitive information leakage occurs when  data passes from trusted to untrusted channels. 
Leading cybersecurity vendors have started to adopt DLP \cite{katzer2013office} \cite{ouellet2011magic} to counter the problem via a detection model that differentiates sensitive information from non-sensitive information. 
Automated data sensitivity detection empowers DLP with the ability to monitor users' actions toward only particular relevant portions of sensitive data rather than tracking all data at all times.
However, many current DLP detection models are incapable of capturing unlabeled sensitive texts. This  raises the following question: How can we detect substantial amounts of unlabeled sensitive texts anywhere within a network without relying on users? Our approach in this paper addresses the issue by partitioning a large text corpus into smaller groups of similar paragraphs wherein multiple similarity-based classification models can be built to predict a paragraph's security label. The challenge that sensitive text detection presents  is that even a few words can possess a great deal of  value. 

The main contribution of our research is a paragraph-level content driven data leak detection method which we evaluate on genuine sensitive data. The datasets that we constructed for our evaluation are derived from U.S. diplomatic cables made public by the WikiLeaks organization. These datasets, which we collectively refer to as the \textit{WikiLeaks Dataset}, contain both sensitive and non-sensitive data along with paragraph-level annotations. 
We are not aware of any other paper that has experimented
on WikiLeaks for both sensitive and non-sensitive data. Unlike most prior research, which has treated automated text security  detection as a binary classification problem, our approach more realistically performs classification across multiple security levels. The paragraph-based classification approach that we employ in this paper is also more practical than document-based classification, especially for large documents and can be trivially converted to document-based classification. Finally, note that our ACESS detection model is built with integration into DLP systems in mind. The remainder of this paper is structured as follows: related work and an overview of DLP are presented in Sec.~\ref{Related} and Sec.~\ref{Secapp}. Sec.~\ref{STSC} details our ACESS approach. Sec.~\ref{WikiLeaks} describes the WikiLeaks dataset. The results of our evaluations are presented in Sec.~\ref{Evaluation}. Conclusions and future work are discussed in Sec.~\ref{Conclusion}.

\section{Related Work}\label{Related}


While the use of statistical analysis to detect sensitive text is relatively recent, several techniques have been proposed.
Katz et al. \cite{katz2014coban} employed the $k$-means algorithm on cosine similarity to cluster all the documents in a corpus regardless of their sensitivity level. 
They then assigned a confidentiality score to each document by calculating the confidential term probability. 
Alneyadi et al. \cite{alneyadi2014semantics} used $L_1$-norms \cite{krause2012taxicab} between $n$-gram category profiles, assigning the document to the category of shortest distance.
Hart et al. \cite{hart2011text} proposed a new training method to overcome the problem of imbalanced data by implementing class-specific classifiers. 
Gomez-Hidalgoy et al. \cite{gomez2010data}  proposed the usage of \textit{named entity recognition} to detect ``sensitive'' tweets.

In each of the aforementioned works, however, artificial ``sensitive documents'' were constructed from publicly available sources like articles, tweets, and forums for performance evaluations. 
None of these works evaluated performance with actual sensitive information.
Therefore, how well performance on the evaluations in these works maps to performance on the actual problem is questionable.

While we designed our ACESS system with DLP in mind, sensitive text classification has other applications too. An obvious forensic extension is to assist in monitoring and tracking down suspicious users post mortem. The benefit of logging  users' activities on classified texts is extended when integrated with Security Information and Event Management (SIEM) solutions.  SIEM collects security related logs to analyze and correlate between security events generated from different sources, such as firewalls and IDS \cite{kent2007guide}.
Additionally, text sensitivity classification can offer support in understanding  insiders' behavior by emulating sensitive documents in honeypots \cite{bringer2012survey}. By replacing sensitive documents with similar non-sensitive documents, a monitoring agent could collect and analyze user behavior.


\section{A Prototypical DLP Framework}\label{Secapp} 
 \begin{figure}[!t]
\centering
{\includegraphics[scale=.35]{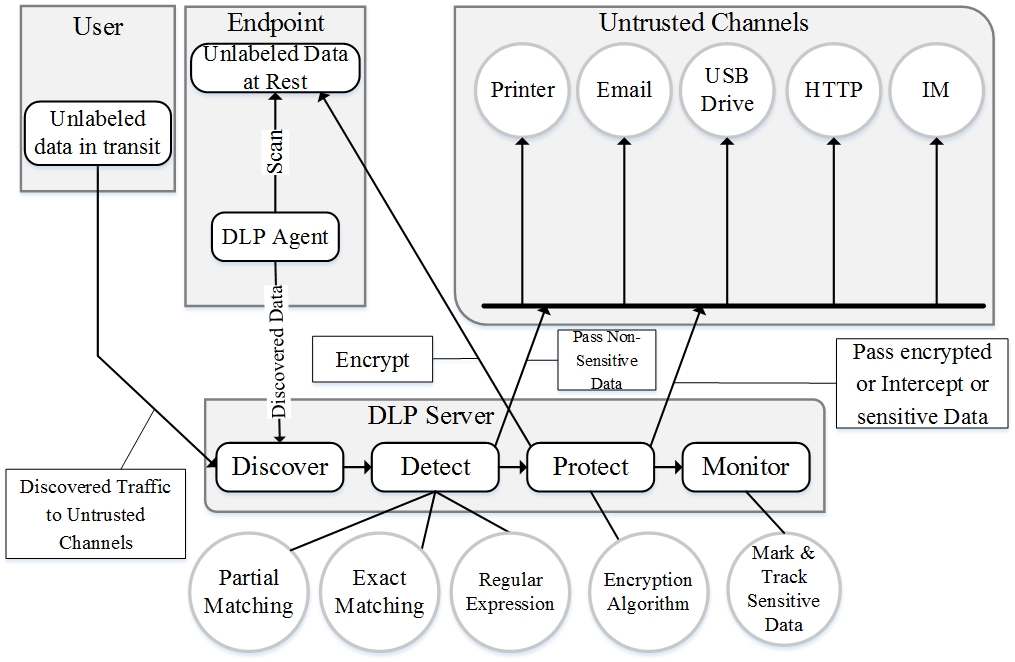}} 
 \caption{\small \textit{A prototypical Data Leak Prevention (DLP) framework: Data can be \textit{discovered} either through scanning end points for unlabeled data or capturing data in transit to untrusted channels. Data is then inspected by the \textit{detection} module to determine its sensitivity level. Common detection techniques include regular expressions, partial matching, and exact matching. Data that is deemed non-sensitive is normally passed to untrusted channels and non-sensitive data at rest is marked as non-sensitive. Detected sensitive data is passed to a \textit{protection} module, which might block the transmission, generate an alert, encrypt the transmission, or encrypt the data at rest. Detected data is then labeled throughout the network by the \textit{monitoring} module which tracks users' future actions with respect to data labeled as sensitive.}}
\label{fig:dlp}
\end{figure}
 DLP  mitigates the threat of sensitive data leakage from insiders by discovering, detecting, protecting, and monitoring sensitive information assets residing on both network and host components (see Fig.~\ref{fig:dlp}). During the \textit{discovery} phase, unlabeled data at rest is located by remotely scanning the targeted device with a scanning agent \cite{mogull2008dlp}. 
This unlabeled data is passed to DLP's \textit{detection} module for sensitivity assessment.  
Data is  then labeled according to its detected sensitivity level. 
Depending on the detected sensitivity level, data is either transmitted to untrusted channels or passed to the \textit{protection} module. The \textit{protection} can protect sensitive data in a variety of different ways and may deal differently with data in transit and data at rest. 
For instance, sensitive data in transit might be encrypted, a security officer might be alerted, or the transmission might be halted.
For data at rest, on the other hand, the DLP's \textit{protection} module might perform encryption and change access rights. 
Finally, the \textit{monitoring} module ensures labeling consistency and tracks users' actions on the sensitive data within the network.
While commercial DLP solutions are not yet standardized, they typically consist of the four modules illustrated in Fig.~\ref{fig:dlp}.  
Our work in this paper focuses on extending \textit{detection} modules by assigning sensitivity labels to textual data. 
In contrast to hard-signature detection approaches which rely on keywords, regular expressions, exact matching and partial matching~\cite{mogull2008dlp}, our approach seeks to infer a more general statistical model to better accommodate previously unseen unstructured texts.


\section{ Learning the Secrets}\label{STSC}

\begin{figure}[tbh]
\centering
{\includegraphics[scale=.43]{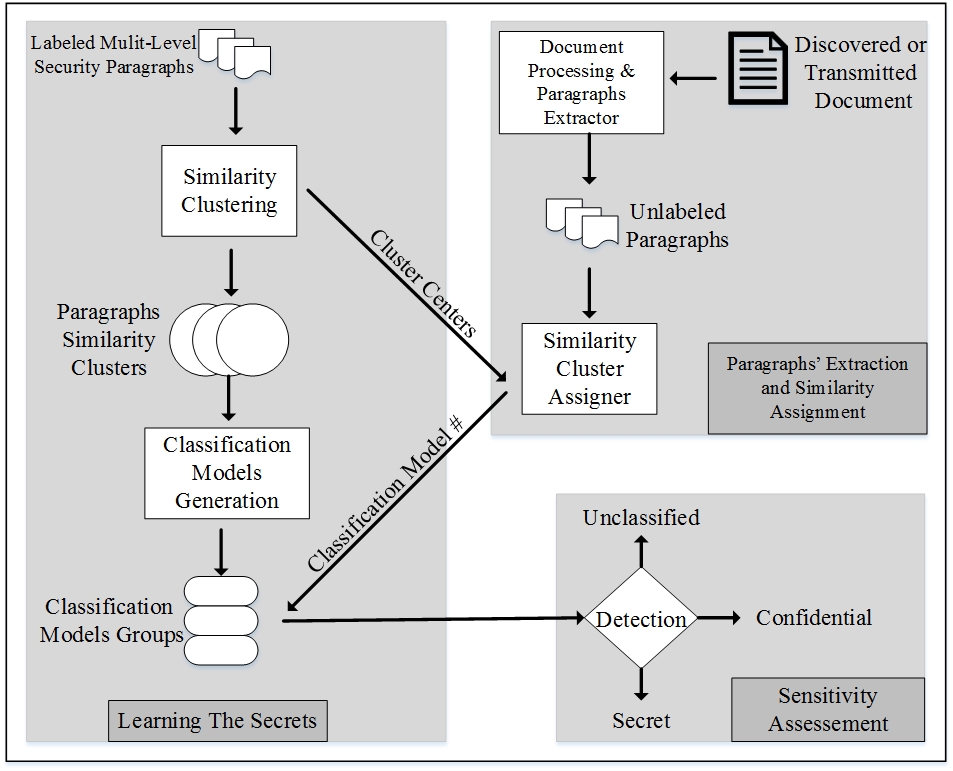}}
 \caption{\small\textit{A schematic representation of the ACESS model. In the first phase, a feature space representation is learnt from real labeled paragraphs for each cluster. Paragraphs are assigned to clusters based on their similarities, and a classification model is generated from each similarity cluster. In the second phase, documents discovered by the DLP's \textit{discovery} module are passed to the closest similarity cluster; the classifier learnt for that cluster is used to assign sensitivity level.}
}
\label{fig:wikileaksdetect}
\end{figure}

 A schematic overview of our ACESS system is presented in Fig.~\ref{fig:wikileaksdetect}. ACESS is a paragraph-based security detection model. Unlike document-based classification, the classifier is built and evaluated on paragraphs independently, regardless of their documents' origins.
Research (e.g., ~\cite{titov2008modeling}) suggests that for a large text corpus such as the WikiLeaks dataset, it can become difficult to differentiate actual security features from spurious similarities brought about by large and diverse paragraphs. 
We hypothesize that a solution to this problem is to partition the dataset into multiple smaller sets of similar paragraphs and build a classifier for each set. 
This approach also has the advantage of producing smaller sets of features to examine for each classifier, allowing for a more principled selection of security features.
One way to do this is to group similar paragraphs together using clustering algorithms. Each similarity cluster and its associated security feature set can then be used to build a \textit{Similarity-Based Classification Model}, wherein each similarity cluster is treated as a small dataset extracted from the original, large dataset.

The optimal number of clusters is indeterminate a priori and impacts the final detection rate. 
One way to determine this number is through cross-validation on the training set, evaluating the number of clusters across a broad range. The Group of Similarity Clusters is a representation of the dataset. Dataset representations, or Group of Similarity Clusters, are independent of one another. Classification Model Group is the set of similarity-based classification models generated
from the same Group of Similarity Clusters. Finally, the Optimal
Classification Model Group is the Classification Model Group that achieves the highest accumulative F-Measure. Another approach would be to use a held-out validation set.

\subsection{Generation of Collection of Groups of Similarity Clusters }

 ACESS's approach to generating the group of similarity clusters is expressed algorithmically in Alg.\ref{alg:the_alg}, and can be described in three steps:
 
\subsubsection{Preprocessing}\label{preprocessing}
During preprocessing, tokenization is first applied, using white space as a delimiter. 
Punctuation and special characters are then removed, and all characters are converted to lowercase. Finally, stop-words are deleted.

\subsubsection{Indexing and Defining Similarity Features}
Tokens are vectorized and transformed to normalized term frequency-inverse document frequency (TF-IDF) values. 
By thresholding TF-IDF values and running validation a selected subset of the TF-IDF features can be used to define a similarity space for clustering paragraphs.

\subsubsection{Group of Similarity Clusters}
The number of produced clusters  varies depending on  the size of the dataset. In this approach, we assumed that there is one cluster for every hundred paragraphs.
Several algorithms can be used for clustering, but ACESS uses the $k$-means algorithm \cite{hartigan1979algorithm}. 
The number of clusters ($k$) is selected via validation and based on the criterion that there be at least two security classes in each cluster.
If the latter condition is not met, the number of clusters is decremented until all clusters meet this condition. A collection of groups of similarity clusters is generated with the same set of similarity features. The entire generation process is outlined in Algorithm \ref{alg:the_alg}.
\begin{algorithm}[htb]
 \algsetup{linenosize=\tiny}
  \scriptsize
    \SetKwInOut{Input}{Input}
    \SetKwInOut{Output}{Output}
	\SetKwFunction{func}{Function}
	\SetKwFunction{algo}{Algorithm}
	\SetKwProg{myfun}{Preprocessing}{}{}
	\SetKwProg{myalgo}{K-Means}{}{}

    \myfun{\func{}}{
  \nl  \Input{Labeled Paragraphs}
   \nl \For{each paragraph} {Tokenize Paragraphs\;
   \nl Remove Special Characters\;
    \nl Remove Stop-Words\;
    \nl Convert letters to lower case\;}}
    Vector $f \leftarrow $ All unique features\;
    Transform $f$ to Normalized TF-IDF values\;
    $k \leftarrow$  Maximum allowed \# of similarity clusters\;
    $p \leftarrow$ 0.6\;
    Vector $v \leftarrow p \times f$ (Features)\;
    \myalgo{\algo{}}{
	\nl \Input{($k, v$, Processed paragraphs)}
    \nl \Output{$k$ similarity  clusters} 
    }
    \eIf{each cluster in $k$ has at least 2 classes }
      {
     \Repeat{ $k < 2$  }
        {Generate $k-1$ similarity clusters}
      }
      {
      \eIf{$p \textgreater 0.01$}{
      $ p = p -0.1$\;
      \textbf{goto} 18
      }{
		$k = k -1$\; 
	  \textbf{goto} 17
     
      }
      }
    \caption{Generation of Collection of Groups of Similarity Clusters }
      \label{alg:the_alg}
\end{algorithm}

\subsection{Similarity-Based Classification Models} 
\begin{figure}[tbh]
\centering
{\includegraphics[scale=.58]{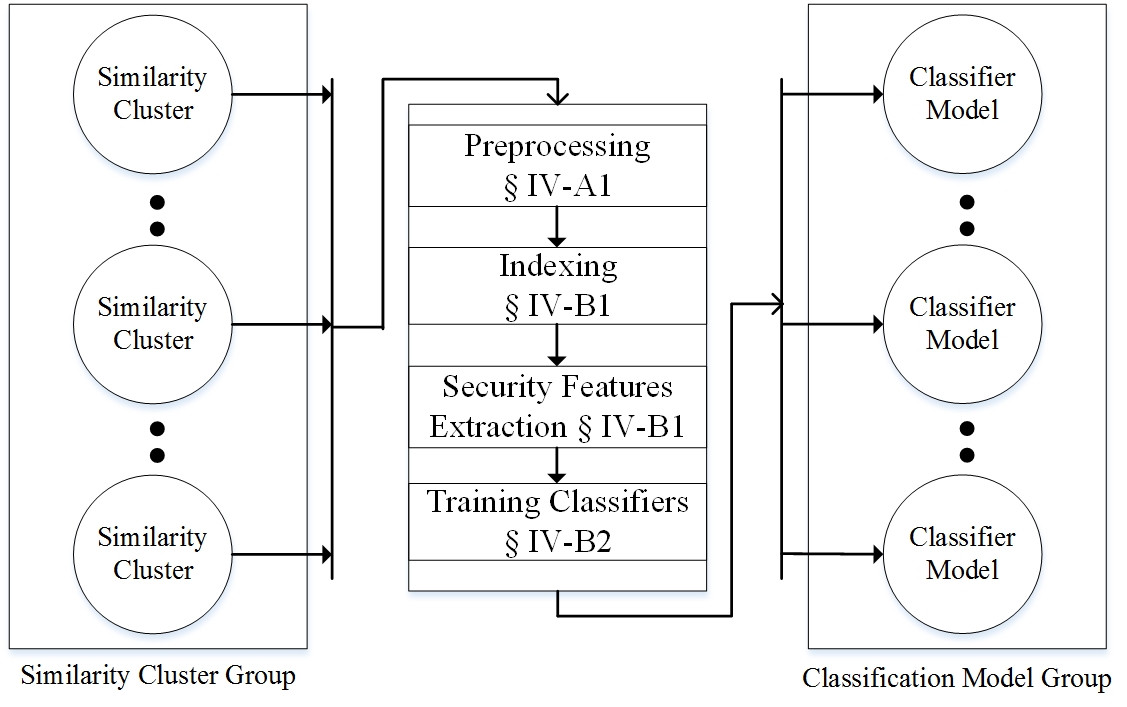}}
 \caption{\small\textit{Similarity based model generation. Left: Pre-defined similarity clusters. Data is converted to a TF-IDF representation via \textit{preprocessing} to index the most similar cluster via the \textit{indexing} step. Security features, are used to train classifiers. The optimal classifiers/security features are arrived at via cross-validation. At query time, a paragraph is assigned to its nearest cluster in TF-IDF feature space, then the features learnt for that cluster in question are then extracted from the paragraph and passed to the that cluster's corresponding classifier.}
}
\label{fig:img3}
\end{figure}

Fig.~\ref{fig:img3} describes the overall process required to construct similarity-based classification models from similarity clusters.
Each cluster within the group is used to train its own separate classification model.  Instances are preserved in their original form without any preprocessing. The following is performed on every group of similarity clusters   to determine the best set of models.  As in any text classification,   a preprocessing identical to the one illustrated in subsection \ref{preprocessing} is first applied to the paragraphs.

\subsubsection{Indexing and Security Features Selection}
Tokens generated in the preprocessing procedure are vectorized and weighted with non-normalized frequency values. Through exhaustive experimenting, we found that TF-IDF and normalization  do not perform as well as term frequencies. The most effective threshold of selected features varies among similarity clusters; thus, determining a decisive model's security feature threshold  requires an extensive search. We used a correlation feature selection algorithm from the Weka toolkit~\cite{hall2009weka} to arrive at a distinct feature space representation per cluster, ranking each feature's value in terms of Pearson correlation with respect to the targeted class \cite{hall1999correlation}. It should be noted that features are extracted from within a similarity cluster. Similarity features will not hold much weight,  since they are distributed across all the security classes. This will assist in the selection of a new set of features with stronger relation to the classes.
 
\subsubsection{Training Classifiers and Selecting Optimal Classification Model Group}
Models are evaluated through a 10 fold cross-validation technique  to assess the  prediction model performance. Based on the selected set of security features, various similarity-based classification models can be built from a single similarity cluster. A similarity-based classification model  that achieves the highest F-Measure score is chosen as the predictive model. However, there are multiple similarity-based classification models in the classification model group. Some of those models might result in high F-measure, while others do not perform as well. Therefore, true and false positive and negative values of all the similarity-based classification models within a group  were summed to calculate the overall classification model group F-Measure. The similarity-based classification model group with the highest F-measure for all security classes was then selected as the optimal classification model group.

%
%
%

\section{The WikiLeaks Dataset}\label{WikiLeaks}

WikiLeaks is an organization which collects and publicizes leaked sensitive documents.
The organization gained particular notoriety when it published sensitive U.S. diplomatic cables dated between 2003 and February 2010, the largest collection of leaked sensitive documents distributed by the organization to date.
These cables originated from U.S. Embassies and Consulates from around the globe, and are stored on the WikiLeaks web page in static HTML format, mixed and sorted by their dates of creation. 
By reorganizing these documents based on their geographic locations, we were able to create several distinct datasets of related cables -- one per embassy. We then stripped the HTML tags, and categorized each of the paragraphs in terms of three fields: a UID consisting of sender, receiver, and time stamp; the raw text itself; and the classification label. 
We then divided classifications into three categories: \textit{Unclassified}, \textit{Confidential}, and \textit{Secret}.
Due to the sparsity of meta-labels, e.g., C/FORN and C/NOFORN corresponding to whether confidential information should or should not be shown to foreign nationals, we merged the corresponding entries into more general classification categories, e.g., \textit{Confidential}.
In total, we created datasets from the cables of four embassies: Baghdad, London, Berlin, and Damascus.
We collectively refer to these datasets as the \textit{WikiLeaks Dataset}.
Statistics for each of the collected datasets in terms of document and paragraph classifications are shown in Tables~\ref{tab1} and ~\ref{tabP}.
Since classified documents may contain paragraphs labeled across multiple security levels, we classify each document by the highest security label across all paragraphs.

\begin{table}[!h]
\footnotesize\setlength{\tabcolsep}{2.8pt} 
\caption{\small\textit{Document dataset statistics. The number of instances per sensitivity class per dataset are reported along with the number of TF-IDF features selected.}}
\setlength\extrarowheight{5pt}
\renewcommand{\arraystretch}{2}
 \label{tab1}
  \resizebox{\columnwidth}{!}{%
\begin{tabular}{|>{\Huge}c|>{\Huge}c|>{\Huge}c|>{\Huge}c|>{\Huge}c|>{\Huge}c|}
\hline
\centering\textbf{Dataset}   & \centering\textbf{ \#  Documents}  & \centering\textbf{ \# Unclassified} & \centering\textbf{\# Confidential}   & \centering\textbf{ \# Secret}   & \centering\textbf{ \# Features}\cr
\hline Baghdad &  6586 & 1373  & 4069  & 1144  & 49118  \\
\hline London &  1037  & 368   & 538  & 131  & 23995\\
\hline Berlin &  1706  & 719 & 721   & 266  & 27542 \\

\hline Damascus &  1377  & 483   & 764   & 130  & 26077 \\
\hline
\end{tabular}
}
\end{table}
\begin{table}[!h]
\footnotesize\setlength{\tabcolsep}{2.8pt} 
\caption{\small\textit{Paragraph dataset statistics. The number of instances per sensitivity class per dataset are reported.}}
\setlength\extrarowheight{5pt}
\renewcommand{\arraystretch}{2}
 \label{tabP}
  \resizebox{\columnwidth}{!}{%

\begin{tabular}{|>{\Huge}c|>{\Huge}c|>{\Huge}c|>{\Huge}c|>{\Huge}c|}
\hline
\centering\textbf{Dataset}   & \centering\textbf{\# Paragraphs}  & \centering\textbf{\# Unclassified} & \centering\textbf{\# Confidential}   & \centering\textbf{ \# Secret}\cr
\hline Baghdad &  49955 & 12599  & 30497  & 6859  \\
\hline London &  6180  & 2491   & 3003  & 686 \\
\hline Berlin &  9631  & 4317 & 4439   & 875  \\
\hline Damascus &  8355  & 2311   & 5173   & 871 \\
\hline
\end{tabular}
}
\end{table}

Note that the number of \textit{Unclassified} or \textit{Confidential} instances in tables \ref{tab1} and \ref{tabP} is always greater than the number of \textit{Secret} instances; often by an order of magnitude. Whether this is an artifact of the existence of fewer Secret documents in the wild or whether \textit{Secret} documents are better protected and thus harder to leak is an important question to consider when using the \textit{WikiLeaks Dataset} to evaluate a DLP system.

\section{Experimental Evaluation}\label{Evaluation}

 \begin{figure*}[!th]
    \centering
    \subfloat[Baghdad]
{{\includegraphics[width=8cm,height=4cm]{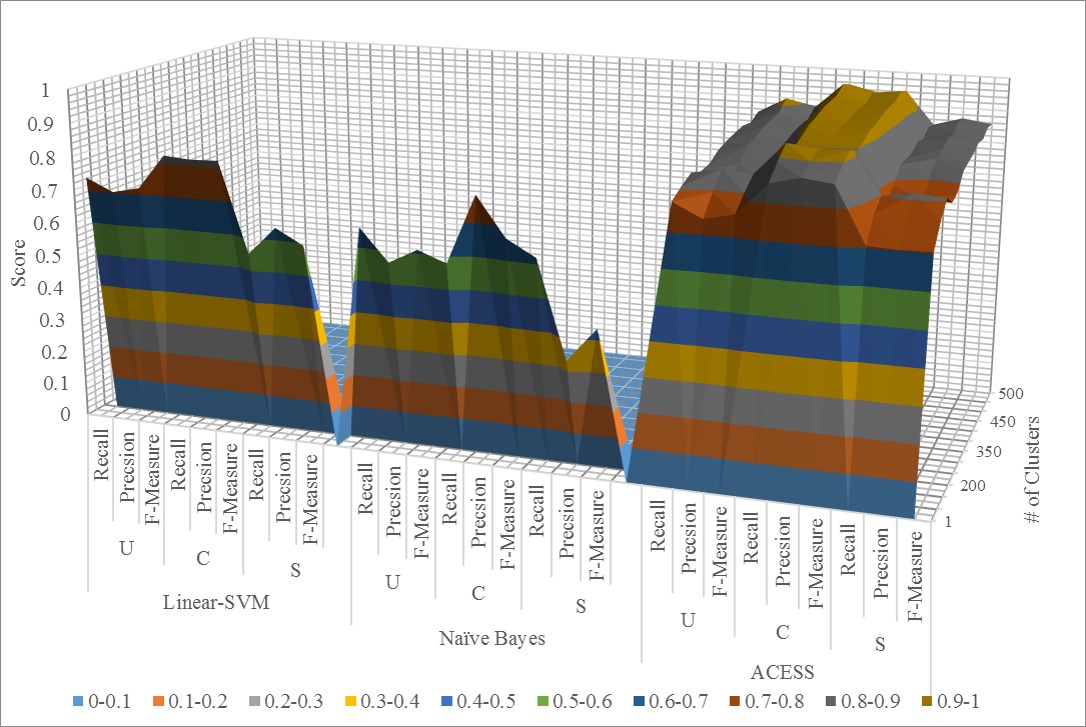} }}%
    \qquad
    \subfloat[Berlin]
{{\includegraphics[width=8cm,height=4cm]{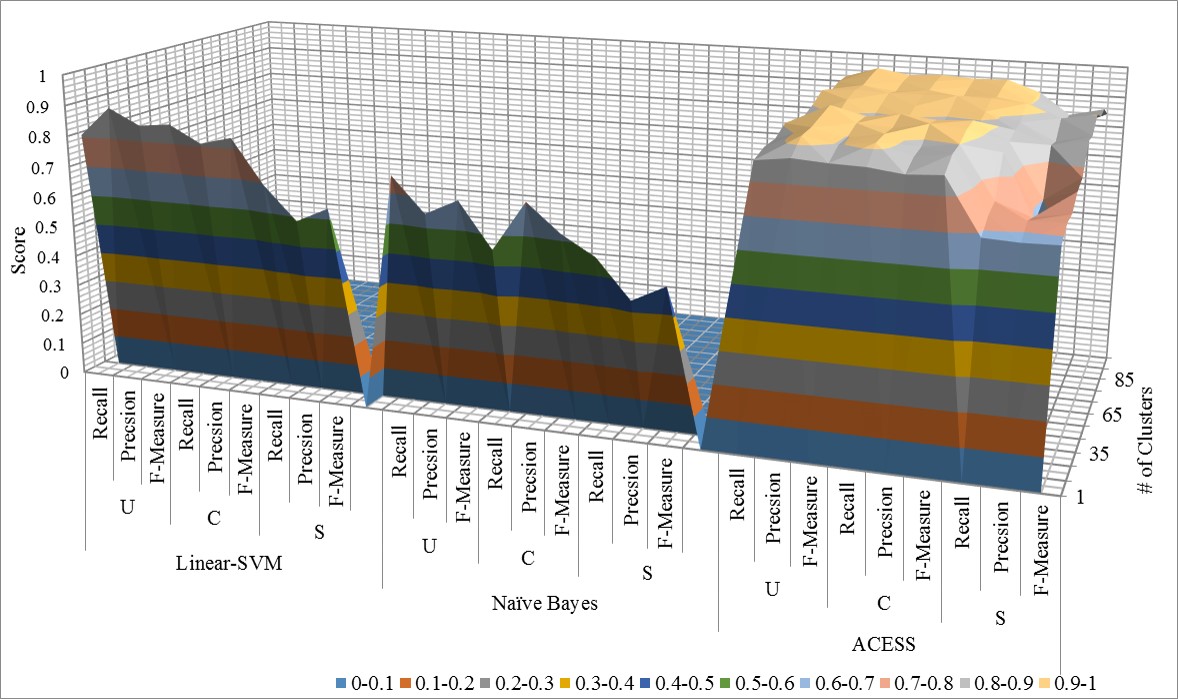} }}%
    
    \subfloat[Damascus]
{{\includegraphics[width=8cm,height=4cm]{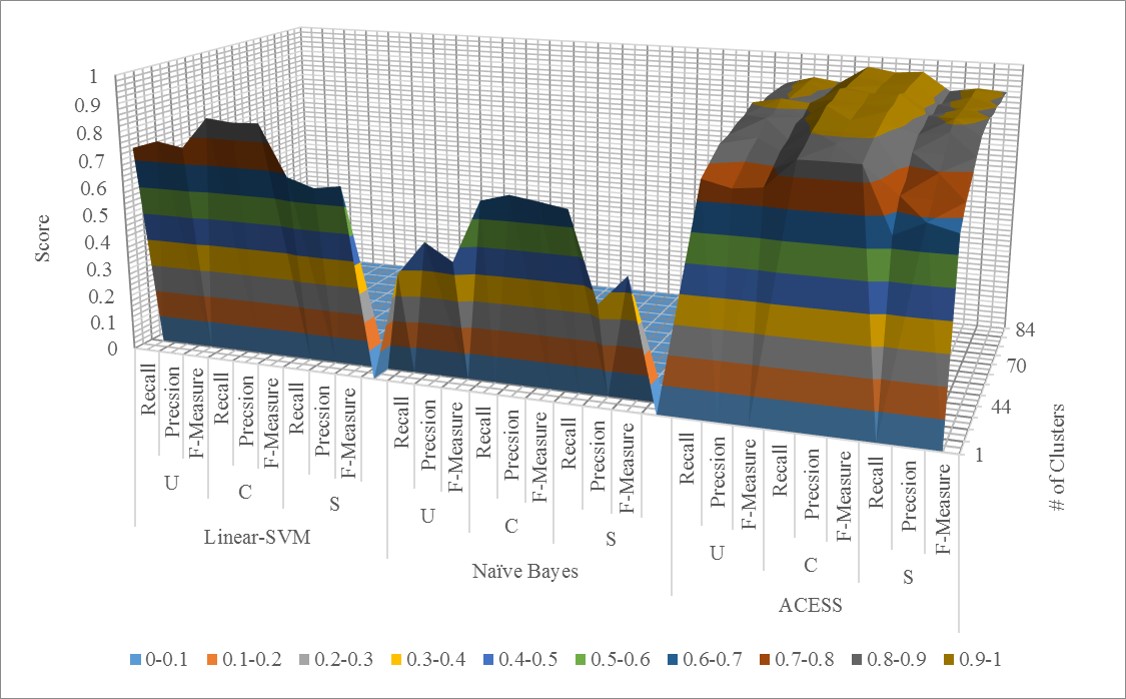} }}%
    \qquad
    \subfloat[London]
{{\includegraphics[width=8cm,height=4cm]{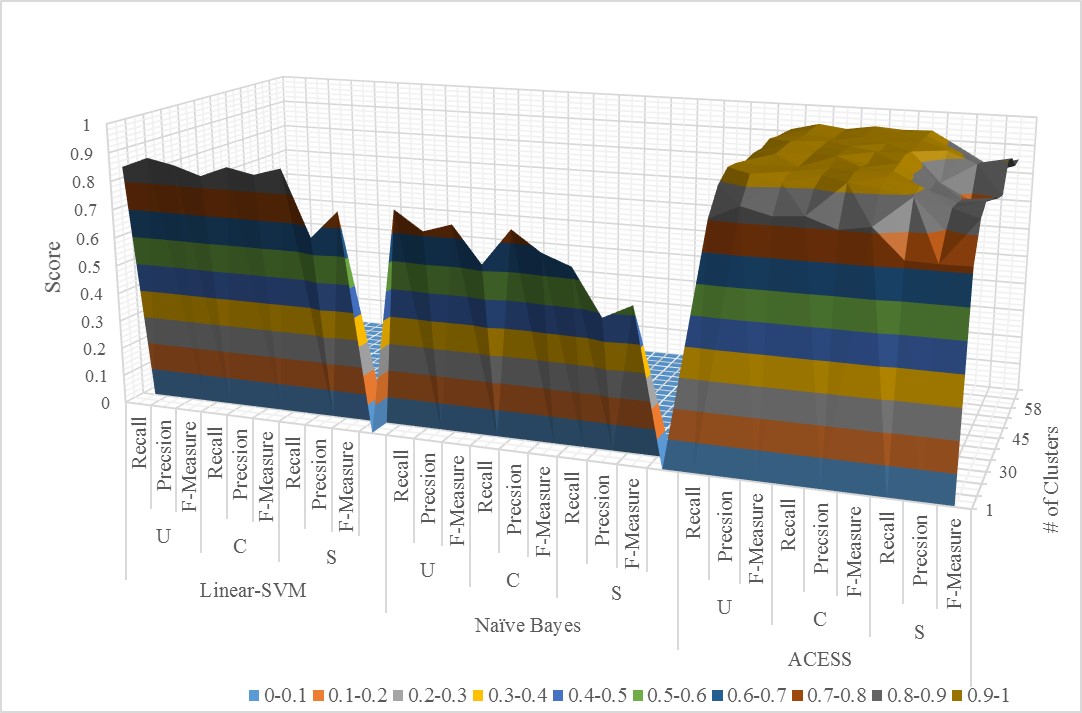} }}%
    \caption{\small\textit{Performance of ACESS and baseline algorithms for each respective embassy of our WikiLeaks dataset is reported in terms of precision, recall, and F1-measure for unclassified (U), confidential (C), and secret (S) classes. Only ACESS has a surface plot, the depth dimension of which is generated by varying the number of clusters ($k$) across a range. For all $k > 1$, ACESS outperforms the baseline models.}}%
    \label{fig:f-measure}%
\end{figure*}

In this section, we evaluate our proposed methodology on the datasets described in Sec.~\ref{WikiLeaks}.
For ease of repetition and implementation, we adopt the widely used practice 
of extracting  TF-IDF features across \textit{each entire dataset}.
 In an actual deployed system, TF-IDF features would be extracted from the training set and possibly background datasets for language priors, e.g., large volumes of text representing common language usage.
Our evaluation, however serves as a feasibility analysis.

\subsection{Evaluation Metrics}
We assume that all security classes -- \textit{Unclassified}, \textit{Confidential}, and \textit{Secret} -- are equally important, which means that raw accuracy numbers alone do not provide an adequate measure due to dataset biases (cf. Tables~\ref{tab1} and ~\ref{tabP}).
Rather, we turn to precision, recall, and F1-measure per-class.
Where $TP$, $FP$, and $FN$ are the numbers of \textit{true positives}, \textit{false positives}, and \textit{false negatives} respectively, these measures are defined as
\[	
	   \textstyle	
	Precision = \frac{TP}{TP+FP}   \qquad 
	Recall = \frac{TP}{TP+FN}
\]  
\[	
   \textstyle
	F1-Measure = \frac{2 \times Precision \times Recall}{Precision + Recall}.
\]

\subsection{Baseline}
Since no previous evaluations have been conducted on the WikiLeaks dataset, we established baseline classification results by testing the multinomial Naive Bayes algorithm and the multi-class 1-vs-1 linear SVM algorithm in a stratified 10-fold cross-validation. Across each fold, we conducted a cross-validated parameter grid search across the training partition, and found that the linear SVM noticeably outperformed Naive Bayes. We also tried several non-linear kernel SVMs, including polynomial and RBF, but found that the respective F1-Measures were at best statistically comparable to the linear SVM \cite{fan2008liblinear} -- a finding which we attribute to the sparsity of the training set and the already high dimensionality of the un-kernelized feature space.

The feature space used for the baseline classifier was obtained by running a correlation algorithm across term frequencies for each embassy's dataset, ranking their frequencies by correlation, and keeping between 0.01\% and 0.6\% of the features. 
For ACESS, we used TF-IDF features for clustering and term frequencies to build classifiers for each cluster. Feature selection was performed independently for each cluster, using correlation with respect to training points in the cluster. Classifiers were built using linear SVMs since they performed the best on the training set.

\subsection{Results}
The results of our evaluation are presented in Fig.~\ref{fig:f-measure}. For the two baseline classifiers, the precision, recall, and F-Measure of each of the three classes are shown. A dimension is added to the plot for the ACESS model, showing the number of clusters ($k$ in $k$-means) selected. For all $k > 1$,    ACESS outperforms both baselines.

Throughout our evaluation, we found that found that for datasets with more instances, more clusters are required to obtain optimal results, for example, Baghdad is the largest of the WikiLeaks datasets that we evaluated and it achieved ``optimal'' performance with 488 clusters. Berlin, Damascus, and London, respectively achieved ``optimal'' performance with 82, 67, and 55 clusters. While the number of clusters cannot be known a priori, it can likely be estimated via cross-validation on the training set. Interestingly, the optimal number of clusters across all datasets was strongly tied to the number of instances in each dataset: Across all four datasets, the optimal cluster count constituted $0.88\pm0.07 \%$ of the data.

\section{Discussion}\label{Discussion}

The fact that ACESS outperforms the baseline algorithms suggests that several classifiers trained in locally derived feature spaces are more discriminative for the type of sensitive data contained in the WikiLeaks dataset than a single monolithic classifier trained in one globally learnt feature space. Equivalently, aggregating all paragraphs under a single TF-IDF representation seems to attenuate the local signal. In our evaluations we were not able to solve this through non-linearity in the classifier alone, as kernelized SVMs did not provide statistically significantly better performance in TF-IDF space.

Instead, ACESS achieves superior classification performance by using global features to direct the query to the right cluster, then perform local, fine-grained classification using that cluster's feature space and decision boundary. 
In this respect, clustering serves as a loose form of \textit{topic modeling}, and an alternative approach, e.g., latent semantic analysis could perhaps be employed.

The surprising result that the optimal number of clusters constitutes roughly the same fraction of each embassy's dataset leads us to hypothesize that there is similarity in topic distribution / sampling across datasets. More analysis is required to verify this hypothesis, but if it holds, several interesting applications could be conducted -- e.g., domain adaptation between embassies and other sensitive contexts could be made more trivial. Domain adaptation for classification of sensitive information is an important topic in its own right, since stakeholders might not trust a DLP developer with their sensitive data but might still wish to use the DLP system.

Notice that in the surface plots in Fig.~\ref{fig:f-measure}, performance noisily tends to increase, peak, then decrease with the number of clusters, corresponding to the transition from an overly-globalized ACESS model to an overly-localized ACESS model. Over-globalization results from too much aggregation in the local feature spaces, while over-localization is a result of a combined under-sampling in the local feature space representation itself and a local decision boundary with too little support. In practice, however, the superior performance that we achieve over a broad range in number of clusters suggests that whether slightly over-specialized or slightly over-globalized, a useful value of $k$ should be trivial to ascertain.

\section{Conclusion}\label{Conclusion}

To our knowledge, the WikiLeaks dataset is the first dataset available to the research community consisting of actual sensitive information.
The dataset is labeled per-paragraph, with multiple levels of sensitivity, and we hope that it will be of use to future researchers.
Our experiments show compelling evidence suggesting that ACESS can enhance the accuracy of generalized machine-learnt DLP detection modules by synthesizing both local and global feature space information. Integrating the ACESS model into DLP Systems ensures labeling consistency through instant detection of sensitive documents.
We have proven that even with an enormous chunk of textual information, it is possible to identify various sensitivity levels within a document. Automating the paragraphs' security classification maximizes the accessibility to unclassified or public texts that exist along with the sensitive ones.



\newpage
\bibliographystyle{IEEEtran}
\bibliography{ISIWikiLeaks}

\begin{thebibliography}{10}
\providecommand{\url}[1]{#1}
\csname url@samestyle\endcsname
\providecommand{\newblock}{\relax}
\providecommand{\bibinfo}[2]{#2}
\providecommand{\BIBentrySTDinterwordspacing}{\spaceskip=0pt\relax}
\providecommand{\BIBentryALTinterwordstretchfactor}{4}
\providecommand{\BIBentryALTinterwordspacing}{\spaceskip=\fontdimen2\font plus
\BIBentryALTinterwordstretchfactor\fontdimen3\font minus
  \fontdimen4\font\relax}
\providecommand{\BIBforeignlanguage}[2]{{%
\expandafter\ifx\csname l@#1\endcsname\relax
\typeout{** WARNING: IEEEtran.bst: No hyphenation pattern has been}%
\typeout{** loaded for the language `#1'. Using the pattern for}%
\typeout{** the default language instead.}%
\else
\language=\csname l@#1\endcsname
\fi
#2}}
\providecommand{\BIBdecl}{\relax}
\BIBdecl

\bibitem{chen2011auditing}
Y.~Chen and D.~Evans, ``Auditing information leakage for distance metrics,'' in
  \emph{Privacy, Security, Risk and Trust (PASSAT) and 2011 IEEE Third
  Inernational Conference on Social Computing (SocialCom), 2011 IEEE Third
  International Conference on}.\hskip 1em plus 0.5em minus 0.4em\relax IEEE,
  2011, pp. 1131--1140.

\bibitem{harel2010m}
A.~Harel, A.~Shabtai, L.~Rokach, and Y.~Elovici, ``M-score: estimating the
  potential damage of data leakage incident by assigning misuseability
  weight,'' in \emph{Proceedings of the 2010 ACM workshop on Insider
  threats}.\hskip 1em plus 0.5em minus 0.4em\relax ACM, 2010, pp. 13--20.

\bibitem{gantz2012digital}
J.~Gantz and D.~Reinsel, ``The digital universe in 2020: Big data, bigger
  digital shadows, and biggest growth in the far east,'' \emph{IDC iView: IDC
  Analyze the future}, vol. 2007, pp. 1--16, 2012.

\bibitem{katzer2013office}
M.~Katzer and D.~Crawford, ``Office 365 compliance and data loss prevention,''
  in \emph{Office 365}.\hskip 1em plus 0.5em minus 0.4em\relax Springer, 2013,
  pp. 429--481.

\bibitem{ouellet2011magic}
E.~Ouellet and R.~McMillan, ``Magic quadrant for content-aware data loss
  prevention,'' \emph{Gartner Group Research Note}, 2011.

\bibitem{katz2014coban}
G.~Katz, Y.~Elovici, and B.~Shapira, ``Coban: A context based model for data
  leakage prevention,'' \emph{Information Sciences}, vol. 262, pp. 137--158,
  2014.

\bibitem{alneyadi2014semantics}
S.~Alneyadi, E.~Sithirasenan, and V.~Muthukkumarasamy, ``A semantics-aware
  classification approach for data leakage prevention,'' in \emph{Information
  Security and Privacy}.\hskip 1em plus 0.5em minus 0.4em\relax Springer, 2014,
  pp. 413--421.

\bibitem{krause2012taxicab}
E.~F. Krause, \emph{Taxicab geometry: An adventure in non-Euclidean
  geometry}.\hskip 1em plus 0.5em minus 0.4em\relax Courier Corporation, 2012.

\bibitem{hart2011text}
M.~Hart, P.~Manadhata, and R.~Johnson, ``Text classification for data loss
  prevention,'' in \emph{Privacy Enhancing Technologies}.\hskip 1em plus 0.5em
  minus 0.4em\relax Springer, 2011, pp. 18--37.

\bibitem{gomez2010data}
J.~M. Gomez-Hidalgo, J.~M. Martin-Abreu, J.~Nieves, I.~Santos, F.~Brezo, and
  P.~G. Bringas, ``Data leak prevention through named entity recognition,'' in
  \emph{Social Computing (SocialCom), 2010 IEEE Second International Conference
  on}.\hskip 1em plus 0.5em minus 0.4em\relax IEEE, 2010, pp. 1129--1134.

\bibitem{kent2007guide}
K.~Kent, ``Guide to computer security log management,'' 2007.

\bibitem{bringer2012survey}
M.~L. Bringer, C.~A. Chelmecki, and H.~Fujinoki, ``A survey: Recent advances
  and future trends in honeypot research,'' \emph{International Journal of
  Computer Network and Information Security}, vol.~4, no.~10, p.~63, 2012.

\bibitem{mogull2008dlp}
R.~Mogull, ``Dlp content discovery: Best practices for stored data discovery
  and protection,'' \emph{USA: Securosis, LLC}, 2008.

\bibitem{titov2008modeling}
I.~Titov and R.~McDonald, ``Modeling online reviews with multi-grain topic
  models,'' in \emph{Proceedings of the 17th international conference on World
  Wide Web}.\hskip 1em plus 0.5em minus 0.4em\relax ACM, 2008, pp. 111--120.

\bibitem{hartigan1979algorithm}
J.~A. Hartigan and M.~A. Wong, ``Algorithm as 136: A k-means clustering
  algorithm,'' \emph{Applied statistics}, pp. 100--108, 1979.

\bibitem{hall2009weka}
M.~Hall, E.~Frank, G.~Holmes, B.~Pfahringer, P.~Reutemann, and I.~H. Witten,
  ``The weka data mining software: an update,'' \emph{ACM SIGKDD explorations
  newsletter}, vol.~11, no.~1, pp. 10--18, 2009.

\bibitem{hall1999correlation}
M.~A. Hall, ``Correlation-based feature selection for machine learning,'' Ph.D.
  dissertation, The University of Waikato, 1999.

\bibitem{fan2008liblinear}
R.-E. Fan, K.-W. Chang, C.-J. Hsieh, X.-R. Wang, and C.-J. Lin, ``Liblinear: A
  library for large linear classification,'' \emph{The Journal of Machine
  Learning Research}, vol.~9, pp. 1871--1874, 2008.

\end{thebibliography}
\end{document}